\documentclass[twocolumn,preprintnumbers,a4paper,
superscriptaddress,floatfix,twoside,prd]{revtex4}

\usepackage{multirow}
\usepackage{bm}
\usepackage{fancyh}
\usepackage{epsfig}
\usepackage{amssymb}
\usepackage{latexsym}
\usepackage{times}
\usepackage{slashed}
\usepackage{amsmath}
\usepackage{amsfonts}
\usepackage{amsbsy}
\usepackage{amscd}
\usepackage{bbm}
\usepackage{graphicx}
\usepackage{epstopdf}
\usepackage{grffile}
\usepackage{upgreek}

\newcommand{\eec}{\end{center}}
\newcommand{\bec}{\begin{center}}

\newcommand{\eem}{\end{matrix}}
\newcommand{\bem}{\begin{matrix}}
\newcommand{\eeq}{\end{equation}}
\newcommand{\beq}{\begin{equation}}
\newcommand{\ba}{\begin{array}}
\newcommand{\ea}{\end{array}}
\newcommand{\bea}{\begin{eqnarray}}
\newcommand{\eea}{\end{eqnarray}}
\newcommand{\baq}{\begin{eqnarray}}
\newcommand{\eaq}{\end{eqnarray}}
\newcommand{\beqs}{\begin{subequations}}
\newcommand{\eeqs}{\end{subequations}}

\newcommand{\ecs}{\end{cases}}
\newcommand{\bcs}{\begin{cases}}

\newcommand\eqs[2]{Eqs.~(\ref{#1}) and (\ref{#2})}

\newcommand{\ftn}{\footnotesize}

\newcommand{\TeV}{{\mbox{\rm TeV}}}
\newcommand{\MeV}{{\mbox{\rm MeV}}}
\newcommand{\GeV}{{\mbox{\rm GeV}}}
\newcommand{\keV}{{\mbox{\rm keV}}}

\newcommand{\ZeV}{{\mbox{\rm ZeV}}}
\newcommand{\EeV}{{\mbox{\rm EeV}}}
\newcommand{\PeV}{{\mbox{\rm PeV}}}

\newcommand{\sFig}[2]{Fig.~\ref{#1}-${\sf\small ({#2})}$}

\newcommand{\etal}{{\it et al.\/}}

\def\to{\rightarrow}

\def\lf{\left(}
\def\rg{\right)}

\newcommand\vev[1]{\langle {#1} \rangle}

\newcommand{\sg}{\ensuremath{\sigma}}

\def\mP{m_{\rm P}}

\newcommand{\Eref}[1]{Eq.~(\ref{#1})}
\newcommand{\Sref}[1]{Sec.~\ref{#1}}
\newcommand{\Fref}[1]{Fig.~\ref{#1}}
\newcommand{\Tref}[1]{Table~\ref{#1}}
\newcommand{\cref}[1]{Ref.~\cite{#1}}

\newcommand{\Gff}{\ensuremath{{\Gamma}_{\phi}}}

\newcommand{\wrh}{\ensuremath{w_{\rm rh}}}

\newcommand{\mx}{\ensuremath{m_{\tilde  h}}}

\newcommand{\xx}{\ensuremath{\tilde  h}}

\newcommand{\vtau}{\ensuremath{\uptau}}

\newcommand{\rhofi}{{\ensuremath{\rho_{\phi{\rm i}}}}}
\newcommand{\rhof}{{\ensuremath{\rho_\phi}}}
\newcommand{\rhoR}{{\ensuremath{\rho_{\rm R}}}}
\newcommand{\rhoRi}{{\ensuremath{\rho_{\rm Ri}}}}

\newcommand{\Yx}{\ensuremath{Y_{\tilde  h}}}
\newcommand{\Yxo}{\ensuremath{Y^0_{\tilde  h}}}

\newcommand{\Yxeq}{\ensuremath{Y_{\tilde  h}^{\rm eq}}}

\newcommand{\Trh}{\ensuremath{T_{\rm rh}}}

\newcommand{\Ti}{{\ensuremath{T_{\rm i}}}}

\newcommand{\vtd}{{\ensuremath{\vtau_\star}}}

\newcommand{\vtrh}{\ensuremath{\vtau_{\rm rh}}}

\newcommand{\vtfo}{\ensuremath{\vtau_{\rm fo}}}
\newcommand{\vtnth}{\ensuremath{\vtau_{\rm nth}}}

\newcommand{\dofr}{\ensuremath{g_{\rho*}}}

\newcommand{\dofrh}{\ensuremath{g_{\rho*}^{\rm rh}}}
\newcommand{\dofsh}{\ensuremath{g_{\sf s*}^{\rm rh}}}

\newcommand{\nx}{\ensuremath{n_{\tilde  h}}}

\newcommand{\gx}{\ensuremath{g_{\tilde  h}}}

\newcommand{\cxf}{\ensuremath{C_{{\tilde  h}\phi}}}
\newcommand{\mff}{\ensuremath{m_\phi}}

\newcommand{\sgv}{\ensuremath{\langle \sigma v\rangle}}

\newcommand{\Nx}{\ensuremath{B_{\tilde  h}}}
\newcommand{\Omx}{\ensuremath{\Omega_{\tilde  h} h^2}}
\newcommand{\Omxsc}{\ensuremath{\left.\Omx\right|_{\rm SC}}}
\newcommand{\Omxan}{\ensuremath{\left.\Omx\right|_{\rm AN}}}
\newcommand{\sd}{\ensuremath{{\sf\small s}}}
\newcommand{\hbl}{\ensuremath{{\cal H}}}

\def\Ka{K\"{a}hler potential~}

\def\Kam{K\"{a}hler manifold}

\newcommand{\plk}{{\it Planck}}

\renewcommand{\aa}{${\sf A}_1$}
\newcommand{\ab}{${\sf A}_2$}
\newcommand{\ac}{${\sf A}_3$}
\renewcommand{\ba}{${\sf B}_1$}
\newcommand{\bb}{${\sf B}_2$}
\newcommand{\bc}{${\sf B}_3$}
\newcommand{\ca}{${\sf C}_1$}
\newcommand{\cb}{${\sf C}_2$}
\newcommand{\cc}{${\sf C}_3$}
\newcommand{\da}{${\sf D}_1$}
\newcommand{\db}{${\sf D}_2$}
\newcommand{\dc}{${\sf D}_3$}

\newcommand{\mgr}{\ensuremath{m_{3/2}}}
\newcommand{\mz}{\ensuremath{m_{\dz}}}
\newcommand{\mth}{\ensuremath{m_{\theta}}}

\newcommand{\no}{\ensuremath{N}}
\newcommand{\om}{\ensuremath{\omega}}
\newcommand{\deq}{\ensuremath{r^{\rm eq}_{\xx\star}}}

\newcommand{\dK}{\ensuremath{\Delta K}}
\newcommand{\dz}{\ensuremath{{\delta} z}}

\newcommand{\hd}{{\ensuremath{H_d}}}
\newcommand{\hu}{{\ensuremath{H_u}}}

\newcommand{\mss}{\ensuremath{\widetilde m}}

\newcommand{\lm}{\ensuremath{\lambda_\mu}}
\newcommand{\lh}{\ensuremath{\lambda_{\xx}}}

\newcommand{\Gz}{\ensuremath{{\Gamma}_{\dz}}}

\newcommand{\Gh}{\ensuremath{{\Gamma}_{\dz\to\tilde{h}}}}

\newcommand{\Gr}{\ensuremath{\widetilde{G}}}
\newcommand{\Gm}{\ensuremath{\Gamma_{\phi}}}

\newcommand{\dxx}{\ensuremath{\delta_{\tilde  h}}}

\def\lz{{\sf\small LZ230616}}

\def\bz{{Z^*}}
\def\al{{\alpha}}

\renewcommand{\refname}{{\bf\scshape References}}

\renewcommand{\thesubsection}{{\small\sf\Alph{subsection}}}

\renewenvironment{subequations}{%
\refstepcounter{equation}%
\setcounter{parentequation}{\value{equation}}%
  \setcounter{equation}{0}
  \ignorespaces
}{%
  \setcounter{equation}{\value{parentequation}}%
  \ignorespacesafterend
}

\begin{document}

\title{\bf\scshape Heavy Higgsino Dark Matter With Low Reheating \\ in View of the LZ High-Recoil Event}

\author{\scshape Constantinos Pallis\\ {\it School of Technology,  Aristotle University of
Thessaloniki, Thessaloniki, GR-541 24 GREECE} \\  {\sl e-mail
address: }{\ftn\tt kpallis@auth.gr}}

\begin{abstract}

\noindent {\ftn \bf\scshape Abstract:} We show that a heavy
higgsino $\xx$ with mass $\mx\sim0.1~\PeV$, which can be made
consistent with the recent LUX-ZEPLIN event avoiding constraints
from direct and indirect detection, can account for the present
cold dark matter abundance of the universe if the reheating
temperature is bounded as $\Trh\lesssim595~\GeV$. These values can
be achieved by a possible out-of-equilibrium decay of a modulus
$\phi$ with mass $\mff\gg\mx$ and a non-vanishing branching ratio
into $\xx$'s which plays a decisive role in reducing $\Trh$ below
its upper limit via the non-thermal $\xx$ production. The modulus
$\phi$ can be identified with the scalar component of the
goldstino superfield which is involved in the generation of the
$\mu$ parameter of MSSM (with $\mu\simeq\mx$) via the
Giudice-Masiero mechanism.
\\ \\ {\scriptsize {\sf PACs numbers: 98.80.Cq, 04.50.Kd, 12.60.Jv, 04.65.+e}
%

}

\end{abstract}

\maketitle

\setcounter{page}{1} \pagestyle{fancyplain}

\addtolength{\headheight}{.5cm}

\rhead[\fancyplain{}{ \bf \thepage}]{\fancyplain{}{\it Heavy
Higgsino Dark Matter With Low Reheating in View of the LZ
High-Recoil Event}} \lhead[\fancyplain{}{\it
\leftmark}]{\fancyplain{}{\bf \thepage}} \cfoot{}

\section{Introduction}\label{intro}

The recent announcement of an isolated nuclear-recoil event at
$248~\keV$, {\sf\small LZ230616}, by the \emph{LUX-ZEPLIN}
({\sf\small LZ}) experiment \cite{lzexp} fuelled a plethora of
works
\cite{mauro,colliders,gnmssm,freese,heavy,confrot,fog,tev1,tevg,ketov,otherside,extra,
waqas,anupam,sahu,sdoublet,okada1,okada2,okada3,pq2,pq1,lee,axion,yanagida,barman,freezeIn,
okada4,type2,331,waqas1,composite,pc, khalil, hooper, de, mura,
djouadi, sneut, sheavy,
das1,das2,roy,sesolo,lu,stefano,randal,minimal,dirac,oklahoma,das3,wagner}
aiming to explain this observation -- which, though, does not
establish a firm detection of a \emph{Cold Dark Matter}
({\sf\small CDM}) particle since its significance, after
accounting for look-elsewhere effects, is just $2.6\sigma$.
Focusing on the interpretation of \lz\ via endothermic inelastic
scattering -- cf.~\cref{exoendo,exo,baer,he,boosted} -- a natural
realization is provided by the scattering of a pseudo-Dirac CDM
particle \cite{mauro, freezeIn} coupled to a vector mediator $Z'$.
In such models, \lz\ can be produced if one restricts the mass of
the CDM particle and the small Majorana mass splitting
$\delta_{21}$ between the two nearly degenerate fermionic states.

A highly predictive version of this scenario can be achieved if we
employ the higgsino, $\xx$,
\cite{mauro,freese,confrot,fog,tev1,tevg,ketov,sheavy,heavy,
colliders, 331, gnmssm, wagner} as CDM candidate. It is well known
that $\xx$ naturally arises as the \emph{Lightest Supersymmetric
Particle} ({\sf\small LSP}) within several variants
\cite{lsp1,lsp2,lsp3, 331, wagner, gnmssm,nagata,delgado} of the
\emph{Minimal Supersymmetric Standard Model} ({\sf\small MSSM}).
Compatibility with the CDM abundance dictated by observations
\cite{plcp, act},
\beq \Omega_{\rm CDM}h^2=0.1179\pm0.0018 \label{omcdm}\eeq
at $95\%$ \emph{confidence level} (c.l.), entails a unique value
\cite{mauro} for the mass of $\xx$, $\mx\simeq1.1~\TeV$, if we
assume a pre-nucleosynthesis evolution within the \emph{Standard
Cosmology} ({\ftn\sf SC}) -- see below. On the other hand,
$\delta_{21}$ can be computed self-consistently with the residual
MSSM spectrum and, to leading order, reads \cite{tevg}
\beq \label{dxx} \dxx=M_Z^2\lf {\sin^2\theta_W\over
M_1}+{\cos^2\theta_W\over M_2}\rg,\eeq
where $\theta_W$ is the Weinberg mixing angle, $M_Z$ is the mass
of $Z$ boson whereas $M_1$ and $M_2$ are the
soft-\emph{Supersymmetry} ({\sf\small SUSY}) breaking masses of
the $U(1)_{\rm Y}$ and $SU(2)_{\rm L}$ gauginos. The requirement
that the predicted recoil rate is reduced to approximately one
accepted LZ event selects \cite{mauro} almost entirely
$\dxx\simeq377~\keV$ which, in turn, can be accommodated for
$2M_1\simeq M_2=28~\PeV$ -- assuming gaugino mass unification.

Although quite promising, such an explanation enfaces a number of
challenges. Namely, $\xx$ would lead to more higher-energy recoil
events, which were not reported by the LZ collaboration
\cite{confrot}. Also, the preferred $\dxx$ requires mostly
\cite{tevg} an ugly hierarchy between $M_1, M_2$ and $\mu$ (as
shown above) and, most importantly, it conflicts with the
solar-capture neutrino constraints
\cite{solar,hoopers,pospelov,confrot1,notsogood,mauro2,heavy}.
Indeed, the $\xx$'s can be captured in the Sun and subsequently
annihilate predominantly into electroweak gauge bosons producing
high-energy neutrinos that may be searched for with neutrino
telescopes. More specifically, the preferred by \lz\ $\dxx$ is
incompatible with the lower bound $\dxx> 566~\keV$
\cite{pospelov}, inferred by the IceCube \cite{icecube} 10 years
data which places stringent limits on high-energy neutrinos from
CDM annihilation in the Sun.

The difficulties above can be overcome \cite{heavy} if we consider
$\mx\sim(0.1-1)~\PeV$ -- for another related proposal see
\cref{yanagida}. Indeed, as pointed out in \cref{heavy}, the upper
bound on $\dxx$ from IceCube \cite{icecube} is much less
restrictive than in the cases with $\mx\sim1~\TeV$ and therefore,
the explanation of \lz\ can be accommodated without troubles. As a
bonus, the expected number of events in the high-energy sideband
remains acceptably low. On the other hand, the aforementioned
range of $\mx$'s yields a relic density $\Omx$ higher than that in
\Eref{omcdm} if we insist on the SC, i.e., if we assume
minimalistically that the decoupling of $\xx$ occurs during the
\emph{radiation dominated} epoch which commences after the
primordial inflation. However, our ignorance about the universal
history before \emph{Big Bang nucleosynthesis} ({\sf\small BBN})
allows for other possibilities \cite{Kam,scnarcadi,scnallax,scn}
too. E.g., the out-of-equilibrium decay of a long-lived massive
modulus $\phi$ can generate an episode of low reheating
\cite{riotto, lr, pamela, gondolo,john, yamaguchi, drees,
dreesemd, wimpbernal, microrh,moroisom} which significantly alters
$\Omx$ \emph{with respect to} ({\sf\small w.r.t}) its value within
SC, $\Omxsc$. Thermal and/or non-\emph{Thermal Production}
({\ftn\sf TP}) mechanisms of $\xx$ can be activated with or
without the achievement of \emph{Chemical Equilibrium} ({\ftn\sf
CE}) \cite{pamela,gondolo} which may decrease the resulting value
of $\Omx$ at the acceptable level of \Eref{omcdm}.


Taking advantage of the fitting of \lz\ in \cref{heavy} we here
adapt our analysis in \cref{lr, pamela} to the case of $\xx$ CDM
and specify a $\phi$-decaying scenario which can reproduce the
largest domain of the required $\Trh$ values. The more accurate
determination of $\Trh$ reveals that its required values are lower
than those found in \cref{heavy} and can become even lower if
$\phi$ produces an average number $\Nx$ of $\xx$'s. The
non-vanishing $\Nx$ values are totally natural and expectable in a
complete framework -- even without a direct  $\phi-\xx$ coupling
-- as stressed in \cref{dreesbr,olivebr}. The presence of $\Nx$
allows for non-TP with and without CE -- cf. \cref{lr, gondolo} --
which decrease the $\Trh$ values required by \Eref{omcdm} to a
level easily achieved by the decay of a typical modulus
\cite{moroi,moduli, baerh} within \emph{Supergravity} ({\sf\small
SUGRA}). Contrary to \cref{sheavy} we do not invoke instantaneous
reheating and the freeze-in paradigm \cite{west}. Moreover, we
specify a particle-physics setting where $\phi$ is represented by
the complex scalar component $\dz$ of the goldstino superfield
which is responsible for the SUSY breaking within SUGRA. We here
adopt a phenomenological model introduced in \cref{susyr,susyrn}
and analyzed further in conjunction with an inflationary stage in
\cref{asfhi, blfhi}. The incarnation of $\phi$ by $\dz$ offers the
opportunity to connect its decay with the generation of the $\mu$
parameter of MSSM employing the Giudice-Masiero mechanism
\cite{masiero}. Given that $\mu\simeq\mx$ in a split-like MSSM
spectrum, the whole picture is clearly economical and rather
predictive since it depends on just three parameters ($\mx, \mz$
and $\Nx$).

Below, we recall the basics for the reheating scenario in
\Sref{rhsc} and review our SUSY-breaking model in \Sref{susyr}.
Then, in \Sref{res}, we determine the parameters which match well
with the explanation of \lz\ via heavy $\xx$, taking into account
a number of constraints listed in \Sref{cons}. We summarize our
conclusions in \Sref{con}.

\section{Reheating Process and $\Omx$} \label{rhsc}

We display here the equations which govern the evolution of the
various energy and number densities involved in our scenario. The
initial form of these equations is displayed in \Sref{rhsc1}
whereas a form more convenient for numerical manipulations is
derived in \Sref{rhsc2}. Approximate results for $\Omx$ are given
in \Sref{rhsc3}.

\subsection{\sc\small\sffamily   Initial Form} \label{rhsc1}

We assume that a weakly coupled modulus $\phi$ decays with a
life-time $\Gff^{-1}$ large enough compared to the thermalization
time and so the kinetic equilibrium can be rapidly established.
The energy densities $\rho_\phi$ of $\phi$ and $\rho_{\rm R}$ of
the produced radiation, and the number density $\nx$ of $\xx$'s
satisfy the following Boltzmann equations -- cf. \cref{lr,
gondolo, pamela}:
\beqs\begin{eqnarray}  &&\hspace*{-.55cm}\dot \rho_\phi+3\hbl
\rho_\phi+\Gm \rho_\phi=0,\label{rf}\\
&& \hspace*{-.55cm} \dot\rho_{\rm R}+4\hbl \rho_{\rm
R}-\Gm\rho_\phi-2\mx\sgv \left( n_{\tilde  h}^2 -
n_{\tilde  h}^{\rm eq2}\right)=0,~~~ \label{rR}\\
&& \hspace*{-.55cm} \dot n_{\tilde  h}+3\hbl n_{\tilde  h}+\sgv
\left(n_{\tilde  h}^2 - n_{\tilde  h}^{\rm eq2}\right)-\Nx\Gm
\rho_\phi/m_\phi=0.~~~\label{nx}
\end{eqnarray}\eeqs
Here the overdot denotes derivation w.r.t the cosmic time $t$ and
$\Nx$ is the average number of the $\xx$'s produced per $\phi$
decay with $m_\phi\gg \mx$. It is, actually, the product of the
branching ratio of $\phi$ into $\xx$ times the mean multiplicity
\cite{dreesemd} of the produced $\xx$'s. Moreover, $n_{\xx}^{\rm
eq}$ is the equilibrium number density of $\xx$ given by
\bea \nonumber n^{\rm eq}_{\xx}&=&\frac{g_{\xx}}{2\pi^2}m^2T K_2(\mx/T)\\
&\simeq&\begin{cases} {g_{\xx}T^3/\pi^2} \hspace*{3.05cm}
\mbox{for} ~~\mx\ll T; \\ {\gx}
\mx^3\>x^{\frac{3}{2}}\>e^{-\frac{1}{x}}P_2\lf\frac{1}{x}\rg/{(2\pi)^{\frac{3}{2}}}~~\mbox{for}
~~\mx\gg T,\end{cases}\label{neq}\eea
where $x=T/\mx$ and $g_{\xx}=2$ is the number of degrees of
freedom of $\xx$ and $P_n(z)=1+(4n^2-1)/8z$ is obtained by
asymptotically expanding the modified Bessel function of the
second kind of order $n$, $K_2(1/x)$ for $x\gg1$.

Crucial role in the determination of $\Omx$ plays the
thermal-averaged cross section times the velocity of $\xx$, $\sgv$
in \Eref{nx}. In the limit of pure Higgsino, the dominant
annihilation channel is into gauge bosons. Since charged and
neutral Higgsino states are nearly mass degenerate, the inclusion
of the co-annihilation processes is essential \cite{delgado}. We
consider four coannihilating states with equal mass $\mx\simeq\mu$
$(\xx:=\tilde H_1, \tilde H_2, \tilde H_+, \tilde H_-)$. When
$\mx\gg M_Z$, $\sgv$ is well approximated by
\cite{mauro,sheavy,delgado}
\bea \nonumber\sgv&=&\frac1{16}\lf 2\sg_{11}+2\sg_{12}+8\sg_{1+}+2\sg_{+-}+2\sg_{++}\rg\\
&=&\frac{g^4}{512\pi \mx^2}\lf 21 + 3 \tan^2\theta_{\rm W} + 11
\tan^4\theta_{\rm W} \rg,\label{sgv}\eea
where $g$ is the $SU(2)_{\rm L}$ coupling constant and the
notation of \cref{delgado} is applied for the contributions
$\sg_{ij}$ with $i=1,2,+,-$. We verified that in the limit of SC,
the fulfilment of \Eref{omcdm} requires $\mx\simeq1.1~\TeV$
adopting $\sgv$ above -- cf.
\cref{mauro,freese,confrot,fog,tev1,tevg,ketov,heavy}. Therefore,
we consider that the accuracy of our formula is sufficient enough
for our purposes and ignore further computational refinements
including Sommerfeld enhancement \cite{moroisom} and 1-loop
effects \cite{nagata}.

The Hubble parameter $\hbl$ in Eqs.~(\ref{rf}) -- (\ref{nx}) is
given by
\begin{equation} \label{Hini}
\hbl =\left(\rho_\phi +\rho_{\rm R} \right)^{1/2}/{\sqrt{3}\mP},
\end{equation}
whereas the temperature $T$ and the entropy density ${\sf\small
s}$ are found from the relations
\begin{equation} \rho_{\rm R}=\frac{\pi^2}{30}g_{\rho*}
T^4~~\mbox{and}~~\sd=\frac{2\pi^2}{45}g_{\sd*} T^3,
\label{rs}\end{equation}
where $g_{\rho*}$ and $g_{\sd*}$ are the energy and entropy
density relativistic degrees of freedom. The system of
Eqs.~(\ref{rf}) -- (\ref{nx}) is solved under the following
initial conditions:
\beq\hbl_{\rm i}=m_\phi~~\Rightarrow~~\rho_{\phi{\rm
i}}=3\mff^2\mP^2~~\mbox{and}~~ \rhoRi=\rho_{h_{\rm
i}}=0,\label{init} \eeq
where the subscript i is referred to quantities defined at the
commencement of the $\phi$ decay. The results are obtained for a
temperature $T_{\rm f}\ll\Trh$ where $\Trh$ is defined from the
condition
\beq\rho(\Trh)=\rho_{\rm R}(\Trh),\label{rhdef}\eeq
which can be solved analytically and accurately enough with result
\cite{pamela,phi}
\beq \label{Trh} \Trh=
\left({72/5\pi^2\dofrh}\right)^{1/4}\Gff^{1/2}\mP^{1/2},\eeq where
$\dofrh\simeq10.75-100$ counts for the effective number of the
relativistic degrees of freedom at $\Trh$.

The $\xx$ yield (or comoving number density) can be estimated by
the solution of \Eref{nx} via the expression
\beq \Yx=\nx/\sd.\label{Ydef} \eeq
The relic density $\Omx$ of $\xx$ can be found from the well-know
formula
\begin{equation}
\label{omxa} \Omx={m_{\tilde  h} n_{\tilde  h0}\over\rho_{\rm
c0}}h^2= {\sd_0\over\rho_{\rm c0}}h^2 \mx\Yxo= 2.748
\cdot10^{14}\frac{\mx}{\PeV}\Yx,
\end{equation}
where $\rho_{\rm c0}= 8.099\cdot10^{-47}h^2~\GeV^4$ with $h$ is
the rescaled Hubble parameter today, $h = 100 \hbl _0 {\rm
km/Mpc~s} $ and $\sd_0=2.23\cdot10^{-38}~\GeV^3$. From the form of
\Eref{nx} it is clear that the results on $\Omx$ do not depend
separately on $\Nx$ and $\mff$ but on their combination
\cite{pamela, gondolo}
\beq\label{cxf}\cxf= (1~\PeV)\Nx/\mff,\eeq
where units are included just for convenience. However, the type
of $\xx$ production -- see \Sref{res2} -- does depend on each of
the parameters above.

\subsection{\sc\small\sffamily  Reformulation} \label{rhsc2}

The numerical integration of Eqs. (\ref{rf})--(\ref{nx}) is
facilitated by absorbing the dilution terms. To this end, we
define the following dimensionless variables \cite{pamela}
\beq\label{fdef}  f_\phi=\rho_\phi a^{3},~f_{\rm R}=\rho_{\rm R}
a^4~~\mbox{and}~~ f_{\tilde  h}=n_{\tilde  h} a^3,\eeq
where $a$ is the scale factor. Indeed, \Eref{rf} -- (\ref{nx}) can
be reexpressed in terms of the variables above as follows
\beqs \bea  \hspace*{-.3cm} \hbl f_\phi' &=&-\Gm f_\phi \label{ff},\\
\hspace*{-.3cm} \hbl  f'_{\rm R} &=&\Gm f_\phi a+2\mx
\sgv\left(f_{\tilde  h}^2 - f_{\tilde  h}^{\rm eq2}\right)a^{-2},
\label{fR}\\ \hspace*{-.3cm} \hbl f'_{\tilde  h}&=& -\sgv
\left(f_{\tilde  h}^2 - f_{\tilde h}^{\rm
eq2}\right)a^{-3}+\Nx\Gff f_\phi /m_\phi, \label{fx} \eea\eeqs
where prime denotes derivation w.r.t the parameter
\beq \vtau=\ln\left(a/a_{\rm
i}\right)~\Rightarrow~a^\prime=a~~\mbox{and}~~a=a_{\rm i
}e^{\vtau}\label{vtau} \eeq
with $a_{\rm i}$ corresponding to the onset of the $\phi$
oscillations. It does not affect the final results and can be
conveniently adjusted so that the resolution of the system from
$\vtau_{\rm i}=0$  to $\vtau_{\rm f}>\vtrh$ -- see below -- is
numerically stable. The system of Eqs.~(\ref{ff})--(\ref{fx}) can
be solved from $\vtau=0$ up to a final value $\vtau_{\rm f}$, if
we translate the initial conditions in \Eref{init} using
\Eref{fdef}.

\subsection{\sc\small\sffamily   Approximate Results}\label{rhsc3}

We focus on the regime with $T\geq\Trh$ where $\rhof\geq\rho_{\rm
R}$. Therefore, $\hbl$ in \Eref{Hini} assumes the form \cite{phi}
\beq \hbl\simeq \sqrt{\rhof}/\sqrt{3}\mP=\mff
e^{-3\vtau/2}~~\mbox{with}~~\rhof\simeq\rhofi
e^{-3\vtau},\label{Hfa} \eeq
where we take into account \Eref{init}. Upon substitution in
\Eref{fR} we obtain
\beq \label{rRa} \rho_{\rm R}=\rhoRi
\left(e^{-3\vtau/2}-e^{-4\vtau}\right)~~\mbox{with}~~
\rhoRi=\frac{2}{5}(3\rhofi)^\frac12\mP\Gff.\eeq
The above function $\rhoR=\rho_{{\rm R}}(\vtau)$ reaches a maximum
at $\vtau_{{\rm mx}}=0.39$, below which $\rho_{{\rm R}}$  is
dominated by the first term in the parenthesis. Upon substitution
into the leftmost relation in \Eref{rs} we find that, for
$\vtau<\vtrh$, $T$ decreases as
\beq T=\Ti e^{-3\vtau/8}~~\mbox{with}~~\Ti=\lf
{30\rhoRi}/{\dofrh\pi^2}\rg^{1/4}\label{Ttau} \eeq
where $\vtrh$ is found by solving \Eref{rhdef} with result
\cite{phi}
\beq \vtrh\simeq ({2}/{3})\ln({\rhofi}/{\rhoRi}).\label{trh}\eeq

Based on the approximate expressions above, we can proceed to an
analytic computation of $\Omx$. For the range of the parameters
under consideration, we single out two cases:

\paragraph{} \hspace*{-0.6cm}  If $\Nx\simeq0$, the non-thermal $\xx$ production is
suppressed and so, \Eref{fx} can be solved applying the well-known
freeze-out procedure \cite{lr} which divides the $\vtau$ interval
of $\Yx$ evolution into two distinct regimes separated by a value
$\vtau=\vtfo$, such that
\beq \sgv n_{\xx}^{\rm eq}(\vtfo)=\hbl(\vtfo). \label{eqfo}\eeq
Inserting \eqs{neq}{Hfa} for $x<1$ into the condition above we
arrive at the expression
\bea \nonumber
\vtfo&\simeq&\frac1{15}\Bigg(6\ln\sqrt{\frac25}\frac83
\frac{\pi^{5}\dofr^{\rm
rh(1/2)}\mff^{5/3}}{\mP(\gx\sgv)^{8/3}\Trh^2\mx^{4}}\\
&-&\left.40W_{-1}\lf-2\pi\lf\frac{2\dofr^{\rm rh
1/2}\mx}{375\gx\mP\sgv\Trh^2}\rg^{2/5}\rg\rg,~~~\label{tfo}\eea
where $W_{-1}$ is the Lambert function with branch $-1$. As can be
verified numerically, for $\vtau\leq\vtfo$, $f_{\xx}\simeq
f_{\xx}^{\rm eq}$ whereas for $\vtau>\vtfo$, $f_{\xx}\gg
f_{\xx}^{\rm eq}$ and so \Eref{fx} can be solved trivially
performing the relevant integration from $\vtfo$ until $\vtrh$. We
find
\bea \nonumber f_{\xx}(\vtrh)&=&\Big(\lf c_{\rm fo}f^{\rm
eq}_{\xx}(\vtrh)\rg^{-1}-\frac{2a_{\rm i}^{-3}}{\sqrt{3\rhofi}}\\
&&\mP\sgv\lf e^{-3\vtrh/2}-e^{-3\vtfo/2}\rg\Big)^{-1},
\label{fxsol}\eea
where $c_{\rm fo}$ is a constant of order unity, determined by
comparing the exact numerical solution of \Eref{fx} with the
approximate under consideration one. Taking into account
\Eref{Trh} and plugging the result above into \eqs{fdef}{Ydef}  we
end up with
\bea
&&\hspace*{-1.3cm}\Yx(\Trh)=\frac14\frac{\dofrh}{\dofsh}\frac{\Trh}{\mP^{2}\mff^2}\Bigg(\frac{
e^{3\vtfo}}{c_{\rm fo} n_{\xx}^{\rm
eq}(\vtfo)}+\frac{\sgv}{45\mP\mff^2} \nonumber \\
&&\hspace*{.3cm}\lf 30\mff\mP
e^{-\frac32\vtfo}-(10\dofrh)^{\frac12}\pi
\Trh^2\rg\Bigg)^{-1},\label{yansg}\eea
which gives the final result for $\Omx$ after substitution into
\Eref{omxa}. Note that the dependence of $\vtfo$ on $\mff$ in
\Eref{tfo} compensates for that appearing directly in the formula
above and, as it can be verified numerically, $\Yx(\Trh)$ is
independent of the specific $\mff$ value.

\paragraph{}  \hspace*{-0.6cm}  If $\Nx\neq0$, the non-thermal $\xx$ production
can be made sizable for $\vtau \gtrsim\vtau_{\rm nth}$ when the
third term in \Eref{fx} dominates over the others. Performing the
integrations over $\vtau$ we obtain
\beq f_{\xx}(\vtrh)= \rhofi^{1/2}\Gm\Nx
\frac{\mP}{\mff}\frac{2a_{\rm i}^{3}}{\sqrt{3}}\lf
e^{3\vtrh/2}-e^{3\vtau_{\rm nth}/2}\rg.\eeq
Due to the large exponential increase of the expression above for
$\vtau=\vtrh$, the precise value of $\vtau_{\rm nth}$ is not
generically crucial for the final result and can be neglected.
Replacing $\Gff$ with $\Trh$ via \Eref{Trh} we can determine the
corresponding contribution into $\Yx$ which takes the simple form
\beq Y_{\xx}(\Trh)=\frac{f_{\xx}}{a^3\sd}(\vtrh)= \frac{5}{4}
\frac{\dofrh}{\dofsh}\ \Nx\ \frac{\Trh}{\mff}. \label{yangm}\eeq
Up to some numerical prefactors, our result coincides with the one
given in \cref{markos,scnarcadi,dreesbr}.

\paragraph*{} Our present formulae updates the one in \cref{lr,pamela}.
Specific comparisons between the values of $\Omx$ obtained
employing our analytical and numerical treatment are exposed in
\Sref{res}.

\section{Identifying the Decaying Modulus}\label{susyr}

The long-lasting $\phi$-dominated era, discussed above, can be
orchestrated within a specific particle-physics framework if we
identify $\phi$ with the sgoldstino field which arises in various
SUSY-breaking scenaria. We here apply the model established in
\cref{susyr,susyrn} and briefly reviewed below. Namely, in
\Sref{susyr1} we describe the SUGRA set-up, in \Sref{susyr2} we
display the relevant mass spectrum and in \Sref{susyr3} we mention
the decay widths which are relevant to our scheme.

\subsection{\sc\small\sffamily SUGRA Set-up}\label{susyr1}

The SUSY breaking is implemented within SUGRA via the \emph{vacuum
expectation value} ({\ftn\sf v.e.v}) that the complex scalar
component $Z=(z+i\theta)/\sqrt{2}$ of a chiral superfield $Z$ (the
goldstino) develops. In \cref{susyr,susyrn} this is achieved by
selecting the following superpotential $W_{\rm hd}$ and \Ka
$K_{\rm hd}$ for the relevant (hidden) sector of the theory
\beqs\bea && W_{\rm hd} = m\mP^{2-\nu}Z^\nu ~~\mbox{and}~~
\label{wh} \\ && K_{\rm
hd}=\no\mP^2\ln\lf1+\frac{|Z|^2-k^2{Z_-^4}/{\mP^2}}{\no\mP^2}\rg,
\label{kh} \eea\eeqs
where $Z_-=Z-Z^*$ and $m, k$ are positive free parameters. Our
model enjoys an enhanced $R$ symmetry -- under which $W_{\rm hd}$
and $Z^\nu$ carry the same charges. This is mildly violated,
though, by the non-vanishing $k$ which endows $R$ axion with mass.
Also $\nu$ is an exponent which may, in principle, acquire any
real value if $W_{\rm hd}$ is considered as an effective
superpotential valid close to the non-zero of v.e.v of $Z$,
$\vev{Z}$, which is found by minimizing the SUGRA potential. It
lies at the Minkowski vacuum \cite{susyrn}
\beq
\label{vevs}\vev{z}=2\sqrt{2/3}|\nu|\mP~~\mbox{and}~~\vev{\theta}=0,\eeq
which is achieved without tuning imposing the condition
\beq
\no=\frac{4\nu^2}{3-4\nu}~~\mbox{with}~~\frac34<\nu<\frac32~~\mbox{for}~~\no<0.\label{no}
\eeq
For these $\no$ values and $k\sim0$ $K_{\rm hd}$ parameterizes the
$SU(1,1)/U(1)$ hyperbolic \Kam.

The consideration of $W_{\rm hd}$ in \Eref{wh} together with the
well-known superpotential of MSSM -- without the bilinear coupling
between the two Higgs superfields -- assist to explain the
required magnitude of the $\mu$ parameter of MSSM. To avoid the
offending term we assign $R$ charges equal to $2$ for both $H_u$
and $H_d$ whereas all the other superfields of MSSM carry zero $R$
charges -- $H_u$ and $H_d$ are the Higgs superfields coupled to
the up and down quarks respectively. As regards the total $K$,
this includes besides $K_{\rm hd}$ in \Eref{kh} the following
terms -- cf. \cref{masiero, susyrn,asfhi,blfhi,actfhi}
\beq \dK=\lm\lf{\bz^{2\nu}}/{\mP^{2\nu}}\rg\hu\hd\ +\ {\rm
h.c.}+|X_\al|^2,\label{dK}\eeq
where the dimensionless constant $\lm$ is taken real for
simplicity and the left-handed chiral superfields of MSSM denoted
by $Y_\al$ with $\al=1,...,7$, i.e.,
\bea X_\al= {Q}, {L}, {d}^c, {u}^c, {e}^c,
\hd~\mbox{and}~\hu,\nonumber \eea
with the generation indices being suppressed for simplicity.

\subsection{\sc\small\sffamily Particle Spectrum}\label{susyr2}

The particle spectrum of the theory at the vacuum in \Eref{vevs}
includes the gravitino ($\Gr$) which acquires mass \cite{susyr}
\beqs\beq \label{mgr} \mgr\simeq 2^{\nu}3^{-\nu/2}
|\nu|^{\nu}m\omega^{N/2}~~\mbox{with}~~\om=\frac{2(3-2\nu)}{3}\cdot\eeq
It contains also the (canonically normalized) sgoldstino (or $R$
saxion) and the pseudo-sgoldstino (or $R$ axion) with respective
masses
\beq \mz\simeq\frac{3\om}{2\nu}\mgr ~~\mbox{and}~~
\mth\simeq12k\om^{3/2}\mgr. \label{mzth}\eeq\eeqs
Finally, $\dK$ in \Eref{dK} leads to a non-vanishing $\mu$ and a
common soft SUSY-breaking mass parameter $\mss$ which indicatively
represents the mass level of the SUSY partners. Namely, we obtain
\beq |\mu|=
\lm\lf\frac{4\nu^2}{3}\rg^\nu(5-4\nu)\mgr~~\mbox{and}~~
\mss=\mgr.\label{mus}\eeq
The leftmost relation above constraints $\lm$ with given
$\mu\simeq\mx$ (from the astrophysics of \lz) and $\mgr\sim\mz$.

\subsection{\sc\small\sffamily Sgoldstino Decay}\label{susyr3}

The initial condition on $\rho_{\phi\rm i}$ in \Eref{init}
presumes that it dominates the energy density of the universe
$\rho_{\rm i}$ at the onset of $\phi$ oscillations. The
replacement of $\phi$ with $\dz$ assures the validity of such a
domination since the $\dz$ energy density at the onset of its
oscillations, $\rho_{z\rm i}$ is comparable with $\rho_{\rm i}$
due to its large v.e.v in \Eref{vevs}. Namely, we have
\beq \rho_{z\rm
i}\simeq\mz^2\vev{z}^2\sim\mz^2\mP^2~~\mbox{and}~~\rho_{\rm
i}=3\mP^2\hbl_{\rm i}^2\simeq3\mP^2\mz^2.\eeq
Note that similar conclusion is not easily obtained for the
Peccei-Quinn field \cite{heavy} which, normally, acquires a much
lower v.e.v (of order $1~\ZeV$).

The total decay width $\Gz$ of the (canonically normalized)
sgoldstino $\dz$ predominantly includes the contributions from its
decay into $\hu$ and $\hd$ which reads
\beq\Gz\simeq\frac{2^{4\nu-1}}{3^{2\nu-1}}\lm^2\frac{\om^2}{4\pi}
\frac{\mz^3}{\mP^2}\nu^{4\nu}\,.\label{Gz}\eeq
It exhibits the $\mz^3/\mP^2$ dependence as expected for any
typical modulus \cite{baerh} and dominates over other decay
channels -- see \cref{asfhi}. Possible contribution due to the
decay of $\dz$ into $\theta$ can be evaded if we take
$k\simeq0.01$ which renders $\mth\sim\mff$ -- see \Eref{mzth}.
Despite the weakness of the $z$ interactions -- which leads to low
$\Trh$ in \Eref{Trh} triggering, thereby, the notorious cosmic
moduli problem \cite{baerh,moduli} -- $\Trh$ here turns out to be
adequately high thanks to the large enough $\mu=\mx$ and $\mz>\mu$
needed for the explanation of \lz\ as we show in \Sref{res}.

Although $\Nx\neq0$ can be motivated from higher order processes
\cite{dreesbr,olivebr}, we here assume the existence of a
non-renormalizable direct coupling between $z$ and two $\xx$,
suggested in \cref{moroi,baerh}. The relevant decay width is
helicity-suppressed and can be parameterized as
\beq
\Gh=\frac{\lh^2}{4\pi}\frac{\mx^2\mff}{\mP^2}\,.\label{Gh}\eeq
Similar decay channels to other gauginos may be suppressed, taking
lower values for the corresponding coupling constants.

\begin{table*}[!t]
\begin{center}\renewcommand{\arraystretch}{1.4}
{\small\begin{tabular}{|c||c|c|c|c|c|c|c|c|c|c|c|c|} \hline
{\sf BMP}&\aa&\ab&\ac&\ba&\bb&\bc&\ca&\cb&\cc&\da&\db&\dc\\
\cline{1-13}
{\sc Halo Model}&\multicolumn{3}{|c|}{\sc SHM
$544$}&\multicolumn{3}{|c|}{SHM $610$}&\multicolumn{3}{|c|}{\sc
SHM+LMC $0.26$}&\multicolumn{3}{|c|}{\sc SHM+LMC $0.6$} \\
\hline\hline
$\mx/\PeV$&\multicolumn{3}{|c|}{$0.32$}&\multicolumn{3}{|c|}{$0.272$}&
\multicolumn{3}{|c|}{$0.102$}&\multicolumn{3}{|c|}{$0.25$}\\
$\Omxsc$&\multicolumn{3}{|c|}{$4107$}&\multicolumn{3}{|c|}{$2992$}&
\multicolumn{3}{|c|}{$663$}&\multicolumn{3}{|c|}{$3809$}\\\hline
\hline
\multicolumn{13}{|c|}{\sc Input Parameters of the Cosmological
Scenario}\\\hline\hline
$\Gamma_\phi/{\rm meV}$&$ 1.2$
&$3.5\cdot10^{-2}$&$2.9\cdot10^{-6}$&$1$&$6.5\cdot10^{-2}$
&$1.7\cdot10^{-8} $&$0.35$&$6.5\cdot10^{-3}$&$1.7\cdot10^{-8}$&0.93&$2.6\cdot10^{-2} $&$2.6\cdot10^{-7}$\\
$\cxf/10^{-9}$&$0$&$1.2\cdot10^{-2}$&$1.2$&$0$&$2\cdot10^{-2}$&
$1.9$&$0$&$8.8\cdot10^{-2}$&$35$&$0$&$1.8\cdot10^{-2}$&$5.1$\\
\hline\hline
\multicolumn{13}{|c|}{\sc Output Parameters of the Cosmological
Scenario}\\\hline\hline
%
%
%
$\Trh/\GeV$&$595$&$101$&$1$&$542$&$70$&$0.76$&$322$&$44.5$&$0.11$&$521$&$87.4$&$0.32$\\
%
%
%
$\Omxan$&$0.1$&$0.13$&$0.14$&$0.1$&$0.13$&$0.13$&$0.1$&$0.13$&$0.14$&$0.1$&$0.13$&$0.14$\\\hline
\multicolumn{13}{|c|}{\sc Particle-Model Parameters}\\\hline\hline
$m/\EeV$&&$8.2\cdot10^3$&$0.82$&&$5.4\cdot10^3$&$0.54$&&$1.5\cdot10^4$&$0.04$&&$10^4$&$0.1$\\
$\mz/\EeV$&&$2.1\cdot10^4$&$2.1$&&$1.4\cdot10^4$&$1.4$&&$4\cdot10^4$&$0.1$&&$2.55\cdot10^4$&$0.255$\\
$\mgr/\EeV$&&$1.2\cdot10^4$&$1.2$&&$7.9\cdot10^3$&$0.79$&&$2.3\cdot10^4$&$0.058$&&$1.46\cdot10^4$&$0.14$\\\hline
$\Nx/10^{-4}$&&$2.5$&$0.025$&&$2.8$&$0.026$&&$35$&$0.035$&&$4.6$&$1.3\cdot10^{-3}$ \\
\hline
$\lm/10^{-3}$&&$1.7\cdot10^{-5}$&$0.17$&&$2.2\cdot10^{-5}$&$0.22$&&$2.8\cdot10^{-6}$&$1.1$&&$1.1\cdot10^{-5}$&$1.1$\\
$\lh/10^{-2}$&&$1.2$&$1.1$&&$1.3$&$0.125$&&$4.5$&$0.15$&&$1.7$&$0.09$\\\hline
\end{tabular}}
\end{center}
\caption{\sl\small Input and output parameters of the cosmological
and particle model which render the four best-fit points of
\cref{heavy} compatible with the requirements of \Sref{cons} and
\Eref{id}. For the particle model we fix $\nu=4/5$.}\label{tab1}
\end{table*}\renewcommand{\arraystretch}{1.}

\section{Imposed Requirements}\label{cons}
\setcounter{paragraph}{0}

To be acceptable, the cosmo-particle part of our proposal should be consistent with:\\

\paragraph{} \hspace*{-0.6cm} {The CDM abundance} in the universe dictated by \Eref{omcdm},
\beq \mbox{i.e.,}~~\Omx\simeq0.12\,. \label{omx}\eeq

\paragraph{}  \hspace*{-0.6cm}  {The BBN constraint} which requires
\cite{nsref}:
\beq\Trh\geq4.1~\MeV \label{bbn}\eeq
for $\mff\sim1~\PeV$ and with sizable hadronic branching ratio.
The bound above is a little softened for larger $\mff$ values.

\paragraph{}  \hspace*{-0.6cm}  The avoidance of non-perturbative
and/or in-medium effects which is assured for $\Trh<\mff$
\cite{markos}. Along the same lines, we require $\lm$ and $\lh$ to
be bounded by the perturbative limit $\sqrt{4\pi}=3.5$ i.e., \beq
\lm\leq3.5 ~~\mbox{and}~~\lh\leq3.5. \label{lb}\eeq

\paragraph{}  \hspace*{-0.6cm}  {Constraints on the range of $m_\phi$}
which allow the decay of $\phi$ into a pair of $\xx$'s and the
consistency of $\hbl_{\rm i}$ in \Eref{init} with the
normalization of the power spectrum of the curvature perturbations
generated by the inflaton and the upper bound ($0.03$) on the
tensor-to-scalar ratio \cite{act,pamela}. All in all, we require:
\beq 2\mx\leq m_\phi\lesssim4\cdot10^{13}~{\GeV}. \label{mfb}\eeq

\paragraph{}  \hspace*{-0.6cm}  {The avoidance of the $\Gr$-induced moduli
problem} \cite{koichi,koichi1}, which is related to the possible
additional $\xx$'s production by the late decay of the $\Gr$'s, if
these are produced by the decay of $\phi$.  Following
\cref{baerh}, we block that decay channel imposing \beqs\beq
\mff<2\mgr.\label{mfgr}\eeq Within our particle model this
hierarchy can be easily achieved -- see \Eref{mzth} -- selecting
\beq 3/4<\nu<1. \label{nub}\eeq\eeqs
%

%

\section{Confronting LZ230616}\label{res}

Employing the cosmo-particle framework described in \Sref{rhsc}
and \Sref{susyr} we are ready here to incorporate an explanation
of \lz. In \Sref{res1} we present some \emph{benchmark points}
({\ftn\sf BMPs}) which highlight the salient features of our
proposal, in \Sref{res2} we analyze the $\xx$ production
mechanisms encountered in our set-up and in \Sref{res3} we
delineate the allowed space of parameters.

\subsection{\sc\small\sffamily Benchmark Points}\label{res1}

For the connection with the astrophysics of \lz\ we take into
account \cref{heavy}, according to which $\xx$ with mass
$\mx\sim(0.1-1)~\PeV$ and $\dxx\sim(365-447)~\keV$ suits well with
both \lz\ and IceCube data. The outputs of that study are four
best-fit points -- named here ${\sf A}, {\sf B}, {\sf C}$ and
${\sf D}$ -- which are accumulated in \Tref{tab1} and guide us to
the selection of the parameters of our cosmo-particle set-up. More
specifically, the relevant astrophysical study considers the
\emph{Standard Halo Model} ({\sf\small SHM}) with escape
velocities $v_{\rm esc} = 544~{\rm km/s}$ and $610~{\rm km/s}$.
These variants of SHM are denoted as SHM $544$ and SHM $610$ in
\Tref{tab1} respectively. SHM can be supplemented with a
high-velocity component from the \emph{Large Magellanic Cloud}
({\sf\small  LMC}) with mass fractions $0.26\%$ and $0.6\%$. The
resulting models are abbreviated as SHM+LMC $0.26$ and SHM+LMC
$0.26$ in \Tref{tab1}. Note that for each point we display only
$\mx$ since the corresponding $\dxx$ values are irrelevant for our
analysis. These heavy $\mx$ values proposed in \cref{heavy} yield
a value for $\Omx$ within the SC, $\Omxsc$, much larger than the
expectations -- cf. \Eref{omcdm} -- as shown in the fourth line of
\Tref{tab1}.

In the following four lines of \Tref{tab1} we present the input
and output parameters of our cosmological scenario which assures
acceptable $\Omx$, i.e., numerically consistent with \Eref{omx}.
Therefore, the achieved decrease of $\Omx$, compared to $\Omxsc$,
is almost $3$ orders of magnitude in all cases. For each best-fit
point of \cref{heavy} we propose three BMPs -- with indices 1, 2
and 3 --, from which the first one uses $\Nx=0$, the second one
considers $\mff$ close to its upper bound in \Eref{mfb} (in BMP
\cb\ this bound in saturated) whereas the last one employs $\mff$
(and so $\mgr$ too) of the order of the $M_1$ and $M_2$ values
required for the derivation of the correct $\dxx$ -- see
\Eref{dxx}.  In the cases with $\Nx=0$ we obtain the maximal
possible $\Trh$ value since the necessary reduction of $\Omxsc$ is
the minimal one. Increasing $\Nx$, $\Omx$ increases -- see
\Eref{yangm} -- and so, the dilution has to be reinforced in order
to obtain the correct $\Omx$ in \Eref{omx}. As a consequence,
$\Trh$ must be further reduced. To check the accuracy of our
analytic expressions in \Sref{rhsc3} we also include in
\Tref{tab1} the relevant value of $\Omx$, $\Omxan$, obtained by
applying those formulas. Namely, we use \eqs{yansg}{omxa} for BMPs
\aa, \ba, \ca\ and \da, whereas \eqs{yangm}{omxa} are applied for
the remaining BMPs. We observe small deviations from the numerical
value in \Eref{omx}, especially in the case of freeze-out with
$\Nx\simeq0$ -- where we take $c_{\rm fo}\simeq5$. Similar
systematic discrepancies are unavoidable to such approximations --
cf. \cref{lr, phi} -- and can be attributed to some residual
production of the $\xx$'s for $\vtau>\vtrh$.

\begin{figure*}[!t]\vspace*{-.12in}
\hspace*{-.1cm}\epsfig{file=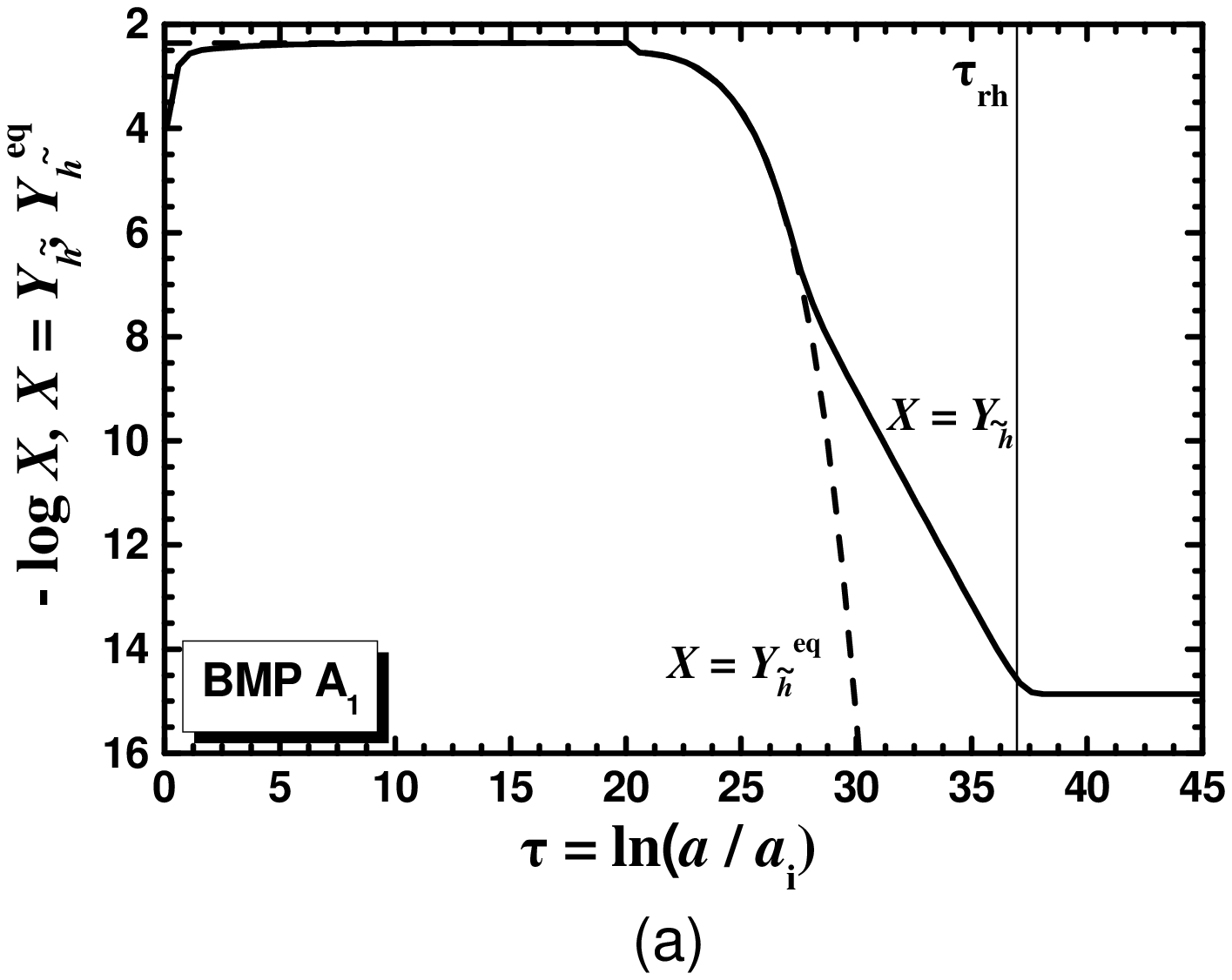,height=2.5in,angle=-90}
\hspace*{-.5cm}\epsfig{file=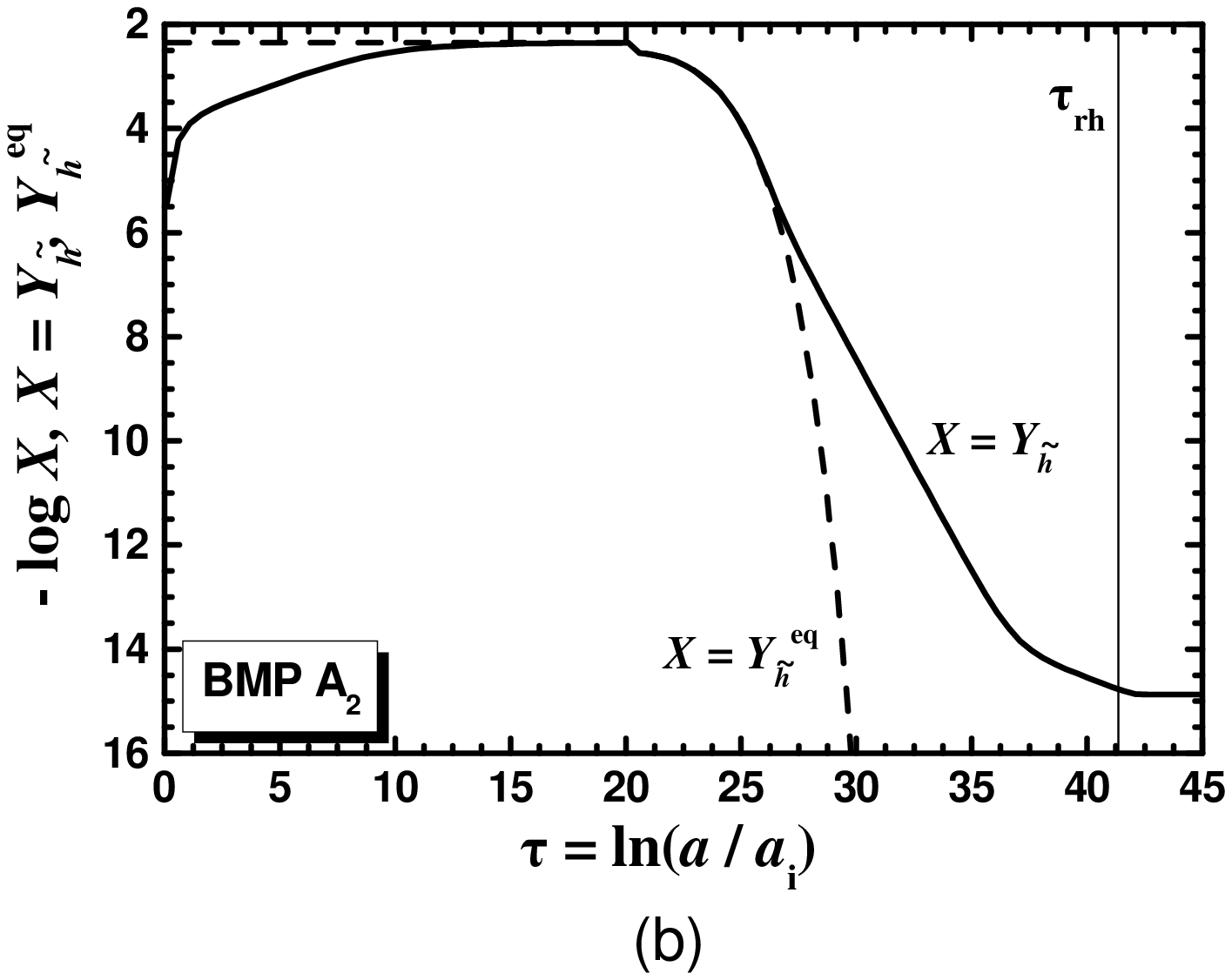,height=2.5in,angle=-90}
\hspace*{-.5cm}\epsfig{file=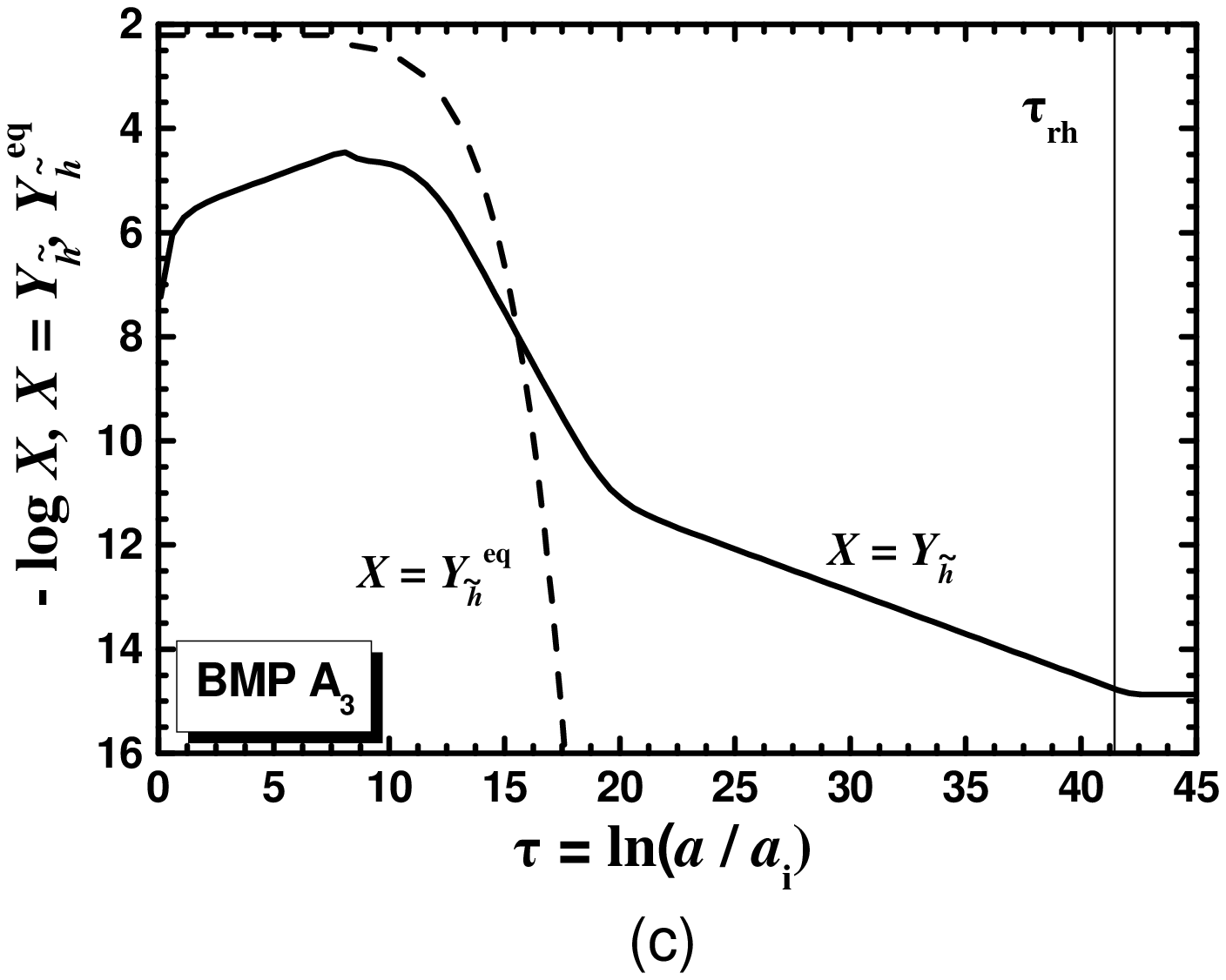,height=2.5in,angle=-90}\hspace*{-.5cm}
\caption{\sl\small Evolution as a function of $\vtau$ of the
quantities: $-\log\Yx$ (solid lines) and $-\log\Yxeq$ (dashed
lines) for BMPs \aa, \ab\ and \ac\ listed in \Tref{tab1} -- see
(a), (b) and (c) panels respectively. The $\vtrh$ value is
depicted by a thin vertical line.}\label{fig1}
\end{figure*}

In the remaining rows of \Tref{tab1} we display parameters of our
particle model in \Sref{susyr} which reproduce the ones of the
cosmological scenario. In other words, we apply the identification
$\phi:=\dz$ which implies
\beq \mff=\mz, \Gff=\Gz
~~\mbox{and}~~\Nx\simeq2\Gh/\Gz,\label{id}\eeq
where the quantities in the right-hand side of the equalities
above are given by \eqs{mzth}{Gz} and the factor of $2$ comes from
the fact that $2$ $\xx$'s are produced per $\dz$ decay with width
$\Gh\ll\Gz$ -- see \Eref{Gh}. Recall that $\mx\simeq\mu$ which
yields $\lm$ through \Eref{mus}. Throughout we fix, consistently
with \Eref{nub}, $\nu=4/5$ which assists to maximize somehow the
derived $\Gz$. For the BMPs \aa, \ba, \ca\ and \da\ we let the
various cells without values since \Eref{id} does not lead to
reasonable values of the particle-model parameters. In particular,
the upper bound of $\mff$ in \Eref{mfb} prevents the increase of
$\Gz$ and $\Trh$ in \eqs{Gz}{Trh} respectively at the
cosmologically required level for fixed $\lm$ (or $\mx$). The
non-zero $\Nx$ values, though, adopted in the other BMPs entail
lower $\Trh$ values which can be easily achieved within our
scheme. The exposed $\Nx$ and $\mff$ combinations are in
accordance with the displayed $\cxf$ values -- see \Eref{cxf} --
in \Tref{tab1}. Also $\mgr=\mss<\mz$ but compatible with
\Eref{mfgr}. As a bonus, for $\mss/3\leq\mu\leq3\mss$ -- see BMPs
with subscript 3 -- our model can become compatible with the mass
of the higgs boson discovered in LHC within high-scale SUSY
\cite{strumia}. For larger hierarchy, though, between $\mu$ and
$\mgr$, this issue requires further investigation. Finally, $\lm$
is derived from \Eref{mus} and $\lh$ from \Eref{Gh} taking into
account the rightmost relation in \Eref{id}. In all cases, $\lm$
and $\lh$ are within the safe ranges of \Eref{lb}.

\subsection{\sc\small\sffamily Production Mechanisms}\label{res2}

To determine more precisely our proposal we specify the
$\xx$-production mechanisms met within it. Recall that the
criteria of the classification of each mechanism are the values of
$\Nx$ and $\deq=\Yx(\vtd)/\Yxeq(\vtd)$ where $\vtd$ is the $\vtau$
value at which the maximum of $\xx$ production -- i.e. of
$\Yx(\vtau)$ -- takes place \cite{riotto,lr}. In particular, for
$\Nx=0$ or $\Nx\neq0$ we obtain TP or non-TP respectively whereas
for $\deq\simeq1$ or $\deq<1$ we obtain $\xx$ production with or
without CE correspondingly. Combining these types of production,
we obtain four categories -- cf. \cref{gondolo, pamela} --, from
which we here do not encounter the one called ``TP without CE'' --
or freeze-in \cite{west} --, since $\sgv$ in \Eref{sgv} is large
enough and ensures CE for $\Nx=0$.

\begin{figure}[!t]
\includegraphics[width=60mm,angle=-90]{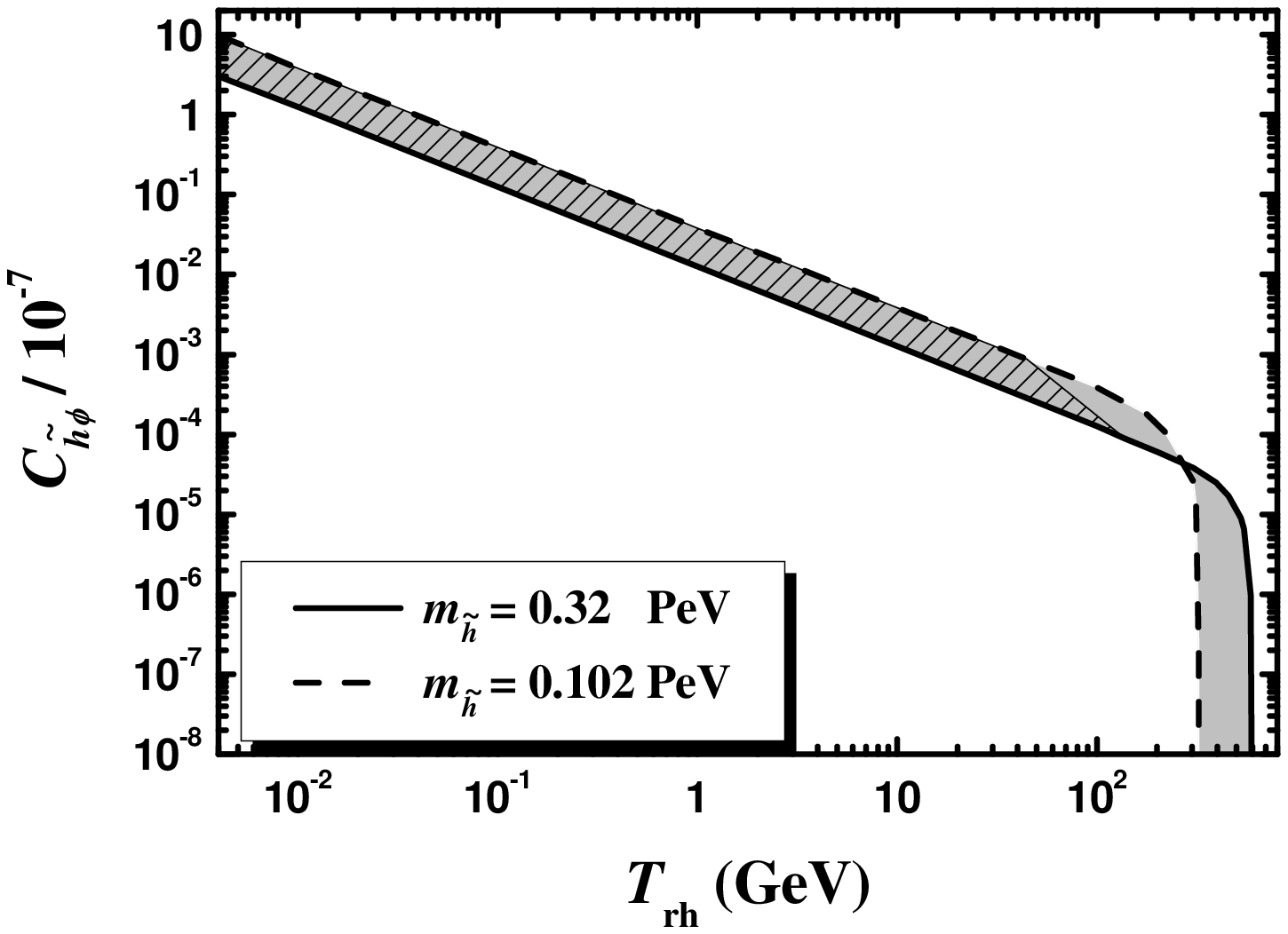}
\hfill\caption{\sl\small  Allowed curves in the $\Trh-\cxf$ plane
for $\mx=0.32~\PeV$ (solid line) and $\mx=0.102~\PeV$ (dashed
line). The cosmologically allowed region is shaded, whereas the
subregion which can be covered by the particle-model parameters
with $\nu=4/5$ is hatched.}\label{fig2}
\end{figure}

The three remaining mechanisms of $\xx$ production are activated
in the cases of the BMPs listed in \Tref{tab1} with indices 1, 2
and 3. For BMP ${\sf A}$ these are illustrated in \sFig{fig1}{a},
{\sf\small (b)} and {\sf\small (c)} where we depict the evolution
as a function of $\vtau$ of $(-\log\Yx)$ (solid lines) and
$(-\log\Yxeq)$ (dashed lines). In all plots $\vtrh$ in \Eref{trh}
is designed by a thin vertical line. More specifically, in
\sFig{fig1}{a} and {\sf\small (b)} we have $\xx$ TP (i.e.,
$\Nx=0$) and $\xx$ nTP (i.e. $\Nx>0$) with CE since in both cases
$\Yx$ reaches $\Yxeq$ for $\vtau=\vtd$, where $\vtd\simeq20.5$
(with $\mff=1~\ZeV$) and $20.2$ for BMP \aa\ and \ab,
respectively. Note that in \sFig{fig1}{b} the activation of the
non-TP is shown by the change of the inclination of the solid line
close to $\vtnth=30.5$.  On the contrary, in \sFig{fig1}{c} no CE
is achieved since $\deq<1$ for $\vtd\simeq8$ whereas the non-TP of
$\xx$ is more clear than in \sFig{fig1}{b} since the difference in
the inclination of the solid line starts for lower $\vtau$,
$\vtnth\simeq20$.

\subsection{\sc\small\sffamily  Allowed Parameters}\label{res3}

We complete our analysis delineating in \Fref{fig2} the overall
allowed range of parameters which can become consistent with the
interpretation of \lz\ via the $\mx$ values in \Tref{tab1}.
Namely, we draw the allowed curves in the $\Trh-\cxf$ plane for
the two limiting best-fit $\mx$ values of \Tref{tab1}, i.e.,
$\mx=0.32~\PeV$ (solid line) and $\mx=0.102~\PeV$ (dashed line).
Obviously the shaded area included between these two curves
represents the overall cosmologically allowed region which
incorporates the uncertainties from the various CDM halo models.
As expected by the findings of \Tref{tab1}, only a part of this
region can be covered consistently with \Eref{id}. To determine
the doubly covered area, we fix $\nu=4/5$ and find the largest
possible $\Trh$ which is localized for the largest $\mff$ in
\Eref{mfb} and represented by a thin line lying in the shaded
area. This line consists the lower boundary of the hatched region
within which reasonable values of the particle-model parameters
allow the reach of cosmologically-compatible results -- for larger
$\nu$'s within \Eref{nub} this region is expected to shrink since
the obtainable $\Trh$'s would be lower. Namely, we find
\beq
\Trh\lesssim139.7~\GeV~~\mbox{and}~~8.8\cdot10^{-4}\lesssim\cxf/10^{-7}\lesssim9.8,\label{resa}\eeq
where the (trivial) lower bound on $\Trh$ in \Eref{bbn} is not
displayed and the upper one is achieved for $\mx=0.32~\PeV$. For
the same $\mx$ we find the lower bound on $\cxf$ whereas its upper
bound is obtained for $\mx=0.102~\PeV$ for $\Trh$ saturating
\Eref{bbn}. Needless to say, finally, the restrictions on $\cxf$
can be easily translated into constraints on $\Nx$ via \Eref{cxf}.
In sharp contrast to \cref{heavy}, the particle-model realization
of our proposal requires $\Nx\neq0$.

\section{Conclusions}\label{con}

Prompted by the recently announced LZ event, \lz, which can be
attributed to endothermic inelastic scattering of heavy -- with
mass $\mx\sim0.1~\PeV$ \cite{heavy} -- higgsino CDM $\xx$ on xenon
nuclei, we investigated how we can obtain the right $\xx$ relic
abundance within a low-reheating scenario. It is worth emphasize
that the employed $\mx$ values evade tight constraints from the
empty sideband of LZ experiment and neutrino telescopes. We
specified three kinds of $\xx$-production mechanisms (thermal or
non-thermal with chemical equilibrium and non-thermal without
chemical equilibrium) constraining the parameters of the
cosmological model ($\Trh$ and $\Nx/\mff$). For the considered
halo models we found $\Trh\lesssim595~\GeV$ and
$\cxf\lesssim9.8\cdot10^{-7}$. In the cases that the non-thermal
$\xx$ production is activated, our scenario can be reconciled with
the decay of a modulus which leads to SUSY breaking and is
responsible for the generation of the parameter $\mu\simeq\mx$ of
MSSM via the Giudice-Masiero mechanism. Namely, the aforementioned
allowed ranges are reduced to the ones in \Eref{resa}.
%

It is gratifying that the majority of the employed here values of
$\Trh$ and $\mz\sim(1-100)~\PeV$ are, also, compatible with F-term
hybrid inflation \cite{asfhi,blfhi,actfhi}. In particular, $m$ in
\Eref{wh} controls the magnitude of the (soft SUSY-breaking)
tadpole parameter ${\rm a}_S$ which plays a crucial role in
fulfilling the observational constraints \cite{act} on the
inflationary observables.

Throughout our investigation we set, as usually, $\wrh=0$ for the
barotropic index during the modulus-decaying era. However, this
may be different especially if $\phi$ is identified with the
inflaton whose the shape of the potential can be substantially
different than the simple quadratic one yielding $\wrh=0$. We
checked that increasing $\wrh$ the required $\Trh$ from \Eref{omx}
can also decrease without to invoke non-zero $\Nx$. We considered
the latter option as more natural and for this reason we did not
analyze the $\wrh>0$ case in our present work. On the other hand,
these $\wrh$ values could help to achieve adequate baryogenesis.
This task is made generally difficult due to the low $\Trh$ values
arising within our scheme. As pointed out recently
\cite{leptolr1,leptolr2}, though, $\wrh>1/3$ may accommodate
sufficient baryogenesis via leptogenesis. Moreover, extensions of
MSSM \cite{allahbau, kanebau} have been also proposed, within
which the late decay of a modulus may generate non-thermally the
required baryon asymmetry of the universe.


\def\ijmp#1#2#3{{\sl Int. Jour. Mod. Phys.}
{\bf #1},~#3~(#2)}
\def\plb#1#2#3{{\sl Phys. Lett. B }{\bf #1}, #3 (#2)}
\def\prl#1#2#3{{\sl Phys. Rev. Lett.}
{\bf #1},~#3~(#2)}
\def\rmp#1#2#3{{Rev. Mod. Phys.}
{\bf #1},~#3~(#2)}
\def\prep#1#2#3{{\sl Phys. Rep. }{\bf #1}, #3 (#2)}
\def\prd#1#2#3{{\sl Phys. Rev. D }{\bf #1}, #3 (#2)}
\def\npb#1#2#3{{\sl Nucl. Phys. }{\bf B#1}, #3 (#2)}
\def\ibid#1#2#3{{\it ibid. }{\bf #1},~#3~(#2)}
\def\cpc#1#2#3{{Comput. Phys. Commun.}
{\bf #1},~#3~(#2)}
\def\astp#1#2#3{{\sl Astropart. Phys.}
{\bf #1},~#3~(#2)}
\def\epjc#1#2#3{{\sl Eur. Phys. J. C}
{\bf #1},~#3~(#2)}
\newcommand\jcap[3]{{\sl JCAP }{\bf #1}, #3 (#2)}
\newcommand\jhep[3]{{\sl JHEP }{\bf #1}, #3 (#2)}
\def\prd#1#2#3{{\sl Phys. Rev. D }{\bf #1}, #3 (#2)}
\def\prdn#1#2#3#4{{\sl Phys. Rev. D }{\bf #1}, no~#4, #3 (#2)}

\newcommand{\hepth}[1]{{\ftn \tt hep-th/#1}}
\newcommand{\hepph}[1]{{\ftn\tt hep-ph/#1}}
\newcommand{\hepex}[1]{{\ftn\tt hep-ex/#1}}
\newcommand{\astroph}[1]{{\ftn\tt astro-ph/#1}}
\newcommand{\arxiv}[1]{{\ftn\tt  arXiv:#1}}


\begin{thebibliography}{999}
 \section*{\refname}




\bibitem{lzexp} D.S. Akerib \etal\ [LUX-ZEPLIN Collaboration], {\it Search for dark
matter particle interactions in an extended nuclear recoil energy
window with the LUX-ZEPLIN (LZ) experiment}, \arxiv{2609.02823}.


\bibitem{mauro} M. Di Mauro, {\it Dark Matter at the Kinematic Edge: Interpreting the
248 keV LZ Nuclear-Recoil Candidate}, \arxiv{ 2609.02608}.


\bibitem{freese} K. Freese and D.P. Theodosopoulos, {\it Higgsino Dark Matter
Interpretation of the LUX-ZEPLIN 248 keV Nuclear-Recoil Event},
\arxiv{2609.01583}.


\bibitem{confrot} N.L. Rodd, B.R. Safdi, T.R. Slatyer and W.L. Xu, {\it Confronting the
Higgsino Interpretation of the LZ Event with the High-Energy
Sideband}, \arxiv{2609.04175}.

\bibitem{fog} J. Fan and M. Reece, {\it Higgsino Above the Sea of Fog}, \arxiv{ 2609.01504}.

\bibitem{tev1}  L. Wu, Y. Zhang and B. Zhu, {\it \TeV\ Higgsino Dark Matter from LZ
Nuclear Recoil to Fermi-LAT Gamma Rays}, \arxiv{ 2609.01590}.

\bibitem{tevg} X. Du and F. Wang, {\it \TeV\ higgsino interpretation of the LZ
high-recoil event with intermediate-scale electroweak gauginos},
\arxiv{2609.04163}.

\bibitem{ketov} D. Frolovsky and S. V. Ketov,
{\it Higgsino dark matter in the Starobinsky supergravity with the
MSSM in light of the LUX-ZEPLIN event}, \arxiv{2609.11241}.

\bibitem{heavy}  K.~Langhoff, {\it Heavy Higgsino Interpretation of the LZ
Event}, \arxiv{2609.09385}.

\bibitem{sheavy} J.~Unwin,
{\it Heavy Higgsino Dark Matter at the Scale of the Vanishing
Higgs Quartic}, \arxiv{2609.28740}.

\bibitem{gnmssm} S. Bisal, J. Cao, and F. Li, {\it Higgsino Dark Matter
Interpretation of the LZ High-Recoil Event in the GNMSSM with
TeV-Scale Gauginos}, \arxiv{2609.07811}.

\bibitem{331} I.~Khan, A.~Muhammad, G.~Mustafa, F.~Atamurotov, A.~Abdujabbarov
and M.~Khan, {\it LZ-Motivated Pseudo-Dirac Higgsinos in the
Supersymmetric 331 Model from the Supersymmetric {SU(6)} GUT
Model}, \arxiv{2609.23691}.


\bibitem{colliders} K. Cheung, S. K. Kang, and R. Kumar, {\it From LUX-ZEPLIN to Colliders:
Probing Higgsino Dark Matter}, \arxiv{2609. 08712}.

\bibitem{wagner}
S.~Roy, P.~Schwaller and C.E.M.~Wagner, {\it Interpreting the
High-Recoil LUX-ZEPLIN Event with Bino-/Singlino-like and Higgsino
Dark Matter}, \arxiv{2609.40234}.



\bibitem{otherside} G. Gu, L. Li, S.-S. Tang and Y. Xu, {\it Inelastic from the Other Side:
Xenon Excitation Signals in Light of the LZ High-Recoil Event},
\arxiv{2609.05291}.

\bibitem{extra} V.S.H. Lee and L. Randall, {\it A Warped Extra Dimensional Candidate
for the LZ 248 keV Event}, \arxiv{2609.09136}.

\bibitem{waqas} W.~Ahmed and G.K.~Leontaris, {\it A Dark-Dimension Origin of
Geometric Inelastic Dark Matter: The LUX-ZEPLIN High-Recoil Event
and Multi-Target Tests}, \arxiv{2609.07138}.


\bibitem{waqas1} W.~Ahmed and G.K.~Leontaris,
{\it Kaluza--Klein Dark-Photon Mediation of Inelastic Dark Matter
at LUX-ZEPLIN}, \arxiv{ 2609.22739}.

\bibitem{randal} V.S.H. Lee and L. Randall, {\it A Warped Extra Dimensional
Candidate for the LZ 248 keV Event}, \arxiv{2609.09136}.


\bibitem{anupam} S. Jeesun and A. Majumdar, {\it Atmospheric neutrino up-scat\-tering
explanation of LZ 2026 excess}, \arxiv{2609.04185}.

\bibitem{sahu} D. Borah, S.K. Sahoo, N. Sahu and S. Sharma, {\it Inelastic
Singlet-Doublet Fermion Dark Matter in light of the 248 keV LZ
event}, \arxiv{2609.07800}.

\bibitem{sdoublet} D. Bandyopadhyay, D. Borah and P. Borah, {\it LZ nuclear recoil
event from inelastic singlet-doublet scalar dark matter},
\arxiv{2609.07451}.


\bibitem{okada1} N. Okada and O. Seto, {\it Inelastic B - L scalar dark matter and the
LUX-ZEPLIN event}, \arxiv{2609.06909}.

\bibitem{okada2} H. Okada and L. Singh, {\it Radiative double inverse seesaw and dark
matter in an alternative gauged U(1)B-L model}, \arxiv{
2609.06494}.

\bibitem{okada3} H. Okada, Y. Shigekami and J.-J. Wu, {\it Can a minimal radiative
seesaw explain the LZ 248 keV event?}, \arxiv{2609.13038}.


\bibitem{pq2} W. Yin, {\it A PQ-Symmetric High-Scale SUSY Interpretation of the
LZ High-Energy Recoil,} \arxiv{2609.01892}.


\bibitem{pq1} L. Visinelli, {\it A Peccei-Quinn Origin for
inelastic electroweak dark matter after LUX-ZEPLIN},
\arxiv{2609.02807}.


\bibitem{lee} H.M. Lee, {\it Inelastic dark matter and baryon flavor symmetry in
light of LUX-ZEPLIN (LZ) experiment}, \arxiv{2609. 06171}.

\bibitem{axion} J. Unwin, {\it Axion Portal Dark Matter and the LUX-ZEPLIN High-Recoil
Event}, \arxiv{2609.04186}.

\bibitem{barman} B. Barman {\it Did LZ see modified gravity?},
\arxiv{2609. 15118}.

\bibitem{freezeIn} D. Cabo-Almeida, F. Costa, D. Feiteira, V. Oliveira,
{\it A Freeze-In Interpretation of the LZ High-Energy Nuclear
Recoil Event}, \arxiv{2609.13130}.


\bibitem{yanagida}  N. Nagata and T.T. Yanagida, {\it Asymmetric
Inelastic Dark Matter and the LUX-ZEPLIN event},
\arxiv{2609.18564}.

\bibitem{okada4} N.~Okada and D.~Raut, {\it Endothermic Z'-Portal Dark Matter:
LZ-LHC Complementarity}, \arxiv{2609.21011}.

\bibitem{type2}
P.K.~Paul, S.K.~Sahoo, N.~Sahu and S.~Sharma, {\it Resurrecting
Electroweak Dark Matter via Type-II Seesaw in light of recent LZ
Event}, \arxiv{2609.22063}.


\bibitem{composite} J.~Sheng and K.~Zhang, {\it The LUX-ZEPLIN Event as Hyperfine
Spectroscopy of Composite Dark Matter}, \arxiv{2609. 23477}.


\bibitem{pc} I.~Khan \etal, {\it Elastic toroidal vector dark matter through a
dark photon in the LUX-ZEPLIN high recoil window}, \arxiv{
2609.25114}.


\bibitem{khalil}
D.~Delepine and S.~Khalil, {\it Model-Independent Sideband
Constraints on Inelastic Dark Matter at the LZ High-Recoil
Candidate}, \arxiv{2609.26698}.

\bibitem{hooper} C.~Gemmell, D.~Hooper and G.~Krnjaic,
{\it A Simple Dark Matter Model to Explain the LZ Event and
Galactic Center Excess}, \arxiv{2609.26570}.


\bibitem{de} B.~De, {\it The 248 keV LZ Recoil: A Possible Hint of Non-SM-Like
Quark Yukawa Couplings with a Scalar-Portal Dark Matter},
\arxiv{2609.23096}.

\bibitem{mura} H.~Murayama and B.~Noether,
{\it Heavy Inert Doublet Reconciles LZ and IceCube},
\arxiv{2609.28819}.

\bibitem{djouadi} G.~Arcadi, M.~di Mauro, A.~Djouadi and F.~Queiroz,
{\it A possible interpretation of the LUX-ZEPLIN recoil event in
the 2HD+a scenario}, \arxiv{2609.17196}.

\bibitem{sneut} K.~Langhoff and H.~Xiao, {\it Sneaky Sneutrino Scattering at LZ},
\arxiv{2609.28616}.


\bibitem{das1}
A.~Das and T.~Nomura, {\it Effect of inelastic scalar dark matter
in hidden $U(1)$ scenario after the LZ nuclear recoil},
\arxiv{2609.15600}.

\bibitem{das2} P.~Das, B.~Karmakar, S.~Mahapatra and P.~K.~Paul,
{\it Inelastic Self-interacting Dark Matter and LUX-ZEPLIN 248 keV
Event in a Dirac Modular Inverse Seesaw}, \arxiv{2609.06825}.

\bibitem{das3} P.~Chanda, B.~Das and S.~Mukherjee, {\it Does the High-Energy
LUX--ZEPLIN Event Suggest Hyperfine Atomic Dark Matter?},
\arxiv{2609.36814}.

\bibitem{lu} Y.~Lou and C.T.~Lu,
{\it Fermionic Dark Matter Absorption and the High-Energy Event in
LUX-ZEPLIN}, \arxiv{2609. 01592}.



\bibitem{roy} U.~Chattopadhyay, D.~Das, R.~Puri and J.~Roy,
{\it Sub-TeV Singlino Dark Matter in light from Sagittarius
A$^\ast$ and LUX-ZEPLIN Nuclear-Recoil Event}, \arxiv{2609.02994}.

\bibitem{sesolo} W.~Kotlarski, K.~Kowalska and E.M.~Sessolo,
{\it GUT-induced FCC signatures of the LUX-ZEPLIN event},
\arxiv{2609. 06750}.


\bibitem{stefano} S.~Palmisano, M.~Tammaro and A.~Tesi,
{\it Inferring dark matter masses and interactions from high
recoil energy events in LUX-ZEPLIN}, \arxiv{2609.15985}.


\bibitem{minimal} S.J.~Lee and T.~Youn, {\it Mixing-suppressed inelastic dark
matter: a minimal model for the LZ 248 keV event}, \arxiv{2609.
09138}.

\bibitem{dirac} P.~Borah, S.~Mahapatra and N.~Nath,
{\it Inelastic Dark Matter at LZ from Radiative Dirac Neutrino
Mass Paradigm}, \arxiv{2609.15027}.

\bibitem{oklahoma} V.~Brdar and D.S.~Chattopadhyay,
{\it A Narrow Neutrino Window for the LZ Event},
\arxiv{2609.30255}.








\bibitem{exoendo} J.B. Dent and J.L. Newstead, {\it Exothermic and Endothermic Inelastic
Dark Matter Interpretations at LZ: Sideband Constraints and Future
Prospects}, \arxiv{2609.04673}.

\bibitem{exo} C.H. de Lima, {\it Exothermic Dark Matter at LZ}, \arxiv{26 09.05204}.

\bibitem{baer} H. Baer and V.~Barger, {\it Exothermic dark matter
and the 248 keV nuclear recoil in LUX-ZEPLIN}, \arxiv{2609.06153}.

\bibitem{he} X.G.~He, X.~Hong and S.~Jeesun,
{\it Hadrophilic inelastic freeze-in dark matter in $q_1-q_2$
gauge extension and the high energy LZ event}, \arxiv{2609.15714}.


\bibitem{boosted} S.~Mahapatra and P.K.~Paul,
{\it Boosted or Inelastic? Discriminating Interpretations of the
LZ 248 keV Event}, \arxiv{2609. 14799}.




\bibitem{lsp1} A.~Delgado and M.~Quir{\'o}s,
{\it Higgsino Dark Matter in the MSSM}, {\sl Phys. Rev. D}
\textbf{103}, no.1, 015024 (2021) [\arxiv{20 08.00954}].

\bibitem{lsp2} K.~Kowalska, L.~Roszkowski, E.~M.~Sessolo and S.~Trojanowski,
{\it Low fine tuning in the MSSM with higgsino dark matter and
unification constraints}, {\sl JHEP} \textbf{04}, 166 (2014)
[\arxiv{1402.1328}].

\bibitem{lsp3}
H.~Baer, V.~Barger, D.~Sengupta and X.~Tata, {\it Is natural
higgsino-only dark matter excluded?}, {\sl Eur. Phys. J. C}
\textbf{78}, no.10, 838 (2018) [\arxiv{1803.11210}].






\bibitem{delgado} N. Arkani-Hamed, A. Delgado and G.F. Giudice, {\it The Well-tempered
neutralino}, {\sl Nucl. Phys. }{\bf B741}, 108 (2006) [\hepph{
0601041}].


\bibitem{nagata} N. Nagata and S. Shirai, {\it Higgsino dark matter in high-scale
supersymmetry}, \jhep{01}{2014}{029} [\arxiv{1410.4549}].






\bibitem{act} T.~Louis \textit{et al.} [ACT Collaboration],
{\it The Atacama Cosmology Telescope: DR6 Power Spectra,
Likelihoods and $\Lambda$CDM Parameters}, \arxiv{2503.14452}.

\bibitem{plcp} N.~Aghanim {\it et al.} [\plk\ Collaboration],
{\it Planck 2018 results. VI. Cosmological parameters }{\sl
Astron. Astrophys. }\textbf{641}, A6 (2020) [\arxiv{1807.06209}].




\bibitem{mauro2} M. Di Mauro and H. Shaikh, {\it Solar Capture Tests of Inelastic Dark
Matter after the LZ High-Recoil Event}, \arxiv{2609. 06760}.

\bibitem{hoopers} T.T.Q. Nguyen, T. Linden, and D. Hooper, {\it Solar Neutrino
Constraints on Inelastic Dark Matter Scattering in Light of Recent
LUX-ZEPLIN Observations}, \arxiv{2609.11833}.


\bibitem{pospelov}  M. Pospelov and H. Ramani, {\it Strong Constraints on Higgsino
Dark Matter from Solar Capture}, \arxiv{2609.02775}.


\bibitem{notsogood}  D. Bose \etal, {\it Not so good $\nu$'s for Higgsino dark matter as LZ
excess: stringent limits from Super-Kamiokande and IceCube},
\arxiv{2609.07807}.

\bibitem{confrot1}  A. Ghosh, I. Chavez and C. Kelso, {\it Confronting the Higgsino
Interpretation of the LZ Event with Astrophysical Uncertainties
and Constraints from Solar Capture}, \arxiv{2609.15321}.


\bibitem{solar} D.~Bose \etal, {\it LUX-ZEPLIN's Stairway to Hea$\nu$en: limits on
elastic scatters of dark matter from solar capture}, \arxiv{
2609.21823}.


\bibitem{icecube} IceCube Collaboration, {\it Search for High-Energy Neutrinos From the
Sun Using Ten Years of IceCube Data}, \arxiv{25 07.08457}.








\bibitem{Kam} M. Kamionkowski and M.S. Turner, {\it Thermal Relics: Do We Know Their
Abundances?} \prd{42}{1990}{3310}.

\bibitem{scn} C.~Pallis, {\it CDM Abundance in non-Standard Cosmologies}, {\tt\ftn hep-ph/0610433}.

\bibitem{scnallax} R. Allahverdi \etal., {\it The First Three Seconds: a Review of
Possible Expansion Histories of the Early Universe,} {\sl Open J.
Astrophys.} {\bf 4} (2021) [\arxiv{2006.16182}].

\bibitem{scnarcadi} G.~Arcadi,
{\it Thermal and non-thermal DM production in non-Standard
Cosmologies: a mini review}, \arxiv{2406.11042}.


\bibitem{john} J.~McDonald, {\it WIMP Densities in Decaying-Particle Dominated Cosmology},
\prd{43}{1991}{1063}.


\bibitem{yamaguchi} T. Nagano and M. Yamaguchi, {\it Late time entropy production and
relic abundances of neutralinos,} {\sl Phys. Lett. B}
\textbf{438}, 267 (1998) [\hepph{9805204}].

\bibitem{riotto} G.F. Giudice, E.W. Kolb and A. Riotto,
{\it Largest temperature of the radiation era and its cosmological
implications}, \prd{64}{2001}{023508} [\hepph{0005123}].


\bibitem{lr} C. Pallis, {\it Massive particle decay
and cold dark matter abundance}, \astp{21}{2004}{689}
[\hepph{0402033}].


\bibitem{gondolo} G.~Gelmini and P.~Gondolo, {\it Neutralino
with the right cold dark matter abundance in (almost) any
supersymmetric model} {\sl Phys. Rev. D} \textbf{74}, 023510
(2006) [\hepph{0602230}].



\bibitem{pamela} C.~Pallis,
{\it Cold Dark Matter in non-Standard Cosmologies, PAMELA, ATIC
and Fermi LAT,} {\sl Nucl. Phys. B} \textbf{831}, 217 (2010)
[\arxiv{0909.3026}].


\bibitem{drees} M.~Drees, H.~Iminniyaz and M.~Kakizaki, {\it Abundance of
cosmological relics in low-temperature scenarios,} {\sl Phys. Rev.
D} \textbf{73}, 123502 (2006) [\hepph{0603165}].


\bibitem{dreesemd} M. Drees and F. Hajkarim, {\it Dark Matter Production in an Early Matter Dominated Era},
{\sl JCAP} {\bf 02}, 057 (2018) [\arxiv{ 1711.05007}].



\bibitem{wimpbernal} N. Bernal and Y. Xu, {\it WIMPs during reheating,} {\sl JCAP}
{\bf 12}, 017 (2022) [\arxiv{2209.07546}].

\bibitem{moroisom} H.~Fukuda, Q.~Li, T.~Moroi and A.~Niki, {\it Non-thermal
production of Higgsino dark matter by late-decaying scalar
fields}, \jhep{06}{2025}{091} [\arxiv{2410.15733}].


\bibitem{microrh} G.~Belanger \etal, {\it micrOMEGAs 7: Beyond standard cosmology},
[\arxiv{2606.06645}].










\bibitem{dreesbr} A.~Banik and M.~Drees,
{\it Non-thermal WIMP production from higher order moduli decay,}
{\sl JCAP} \textbf{12}, 032 (2023) [\arxiv{ 2308.15380}].

\bibitem{olivebr} K. Kaneta, Y. Mambrini and K. A. Olive, {\it Radiative production of
nonthermal dark matter}, {\sl Phys. Rev. D} {\bf 99}, 063508
(2019) [\arxiv{1901.04449}].



\bibitem{moroi} T. Moroi, L. Randall, {\it Wino cold dark matter from anomaly
mediated SUSY breaking}, {\sl Nucl. Phys. }{\bf B570}, 455 (2000)
[\hepph{9906527}].



\bibitem{moduli} G. Kane, K. Sinha and S. Watson, {\it Cosmological Moduli and the
Post-Inflationary Universe: A Critical Review}, {\sl Int. J. Mod.
Phys. D } {\bf 24}, no.~08, 1530022 (2015) [\arxiv{1502.07746}].


\bibitem{baerh} K.J.~Bae, H.~Baer, V.~Barger and R.W.~Deal,
{\it The cosmological moduli problem and naturalness},
\jhep{02}{2022}{138} [\arxiv{2201.06633}].


\bibitem{west} L. J. Hall, K. Jedamzik, J. March-Russell and S. M. West,
{\it Freeze-In Production of FIMP Dark Matter}, {\sl JHEP} {\bf
03}, 080 (2010) [\arxiv{0911.1120}].





\bibitem{susyr} C.~Pallis, {\it Gravity-mediated SUSY breaking, R symmetry,
and hyperbolic K\"ahler geometry,} {\sl Phys.\ Rev.\ D }{\bf 100},
no.~5, 055013 (2019) [\arxiv{1812.10284}].

\bibitem{susyrn} C.~Pallis, {\it SUSY-breaking scenarios with a mildly violated $R$
symmetry,} {\sl Eur. Phys. J. C} \textbf{81}, no.~9, 804 (2021)
[\arxiv{ 2007.06012}].

\bibitem{asfhi} G. Lazarides and C.~Pallis, \textit{Probing the Supersymmetry-Mass Scale With F-term Hybrid
Inflation}, {\sl Phys.\ Rev.\ D {\bf 108}, no. 9, 095055 (2023)}
[\texttt{\ftn arXiv:2309.04848}].

\bibitem{blfhi} C.~Pallis, \textit{PeV-Scale SUSY and Cosmic Strings from F-term Hybrid
Inflation}, {\sl Universe} {\bf 10}, no.~5, 211 (2024)
[\texttt{\ftn arXiv:24 03.09385}].

\bibitem{actfhi} C.~Pallis,
{\it F-Term Hybrid Inflation, Metastable Cosmic Strings and Low
Reheating in View of ACT}, PoS \textbf{CORFU2024}, 206 (2025)
[\arxiv{2504.20273}].


\bibitem{masiero} G.F. Giudice and A. Masiero, {\it A Natural Solution to
the $\mu$ Problem in Supergravity Theories}, {\sl Phys. Lett. B}
{\bf 206}, 480 (1988).

\bibitem{phi} C.~Pallis, {\it Updating GUT-scale pole higgs inflation after ACT DR6,}
\prdn{113}{2026}{015033}{1} [\arxiv{2510. 02083}].

\bibitem{markos} G. Ballesteros, M.A.G. Garcia and M. Pierre, {\it How warm are
non-thermal relics Lyman-$\alpha$ bounds on out-of-equilibrium
dark matter}, {\sl JCAP} {\bf 03}, 101 (2021)
[\arxiv{2011.13458}].



\bibitem{nsref} T.~Hasegawa \etal, {\it MeV-scale reheating temperature and
thermalization of oscillating neutrinos by radiative and hadronic
decays of massive particles,} \jcap{12}{2019}{012} [\arxiv{
1908.10189}].





\bibitem{koichi} M. Endo \etal, {\it Moduli-induced gravitino
problem}, {\sl Phys. Rev. Lett.} {\bf 96}, 211301 (2006)
[\hepph{0602061}].


\bibitem{koichi1} S. Nakamura and M. Yamaguchi, {\it Gravitino production from heavy
moduli decay and cosmological moduli problem revived}, {\sl Phys.
Lett. B} {\bf 638}, 389 (2006) [\hepph{0602081}].





\bibitem{strumia} E.~Bagnaschi, G.F.~Giudice,
P.~Slavich and A.~Strumia, {\it Higgs Mass and Unnatural
Supersymmetry}, \jhep{09}{2014}{092} [\arxiv{1407.4081}].



\bibitem{leptolr1} M.A.G.~Garcia, S.E.~Henrich, W.~Ke and K.A.~Olive,
{\it Leptogenesis and Low Reheating Temperatures},
\arxiv{2607.08663}.


\bibitem{leptolr2} B.~Barman, A.~Basu, D.~Borah and N.~Das,
{\it Leptogenesis with sub-electroweak-scale reheating
temperature}, \arxiv{ 2607.09282}.


\bibitem{allahbau} R.~Allahverdi, B.~Dutta and K.~Sinha, {\it Baryogenesis and
Late-Decaying Moduli,} {\sl Phys. Rev. D }\textbf{82}, 035004
(2010) [\arxiv{ 1005.2804}].

\bibitem{kanebau} G.~Kane and M.W.~Winkler,
{\it Baryogenesis from a Modulus Dominated Universe}, {\sl JCAP}
\textbf{02}, 019 (2020) [\arxiv{ 1909.04705}].

\end{thebibliography}
\end{document}